\documentclass[lettersize,journal]{IEEEtran}
\usepackage{amsmath,amsfonts}
\usepackage{algorithmic}
\usepackage{algorithm}
\usepackage{array}
\usepackage[caption=false,font=normalsize,labelfont=sf,textfont=sf]{subfig}
\usepackage{textcomp}
\usepackage{stfloats}
\usepackage{url}
\usepackage{verbatim}
\usepackage{graphicx}
\usepackage{cite}
\begin{document}

\title{Sensitivity Enhancement of Third-order Nonlinearity \\Measurement in THz Frequency Range}

\author{A. Nabilkova, A. Ismagilov, M. Melnik, M. Zhukova, M. Guselnikov, S. Kozlov, A. Tcypkin

        % <-this % stops a space
\thanks{This paper was produced by the IEEE Publication Technology Group. They are in Piscataway, NJ.}% <-this % stops a space
\thanks{Manuscript received April 19, 2021; revised August 16, 2021. Corresponding author is A. Tcypkin.\\
A. Nabilkova, A. Ismagilov, M. Melnik, M. Zhukova, M. Guselnikov, S. Kozlov, A. Tcypkin
are with Research and Education Center of Photonics and Optical IT, ITMO University (e-mail: tsypkinan@itmo.ru}}

% The paper headers
\markboth{Journal of \LaTeX\ Class Files,~Vol.~14, No.~8, August~2021}%
{Shell \MakeLowercase{\textit{et al.}}: A Sample Article Using IEEEtran.cls for IEEE Journals}

\IEEEpubid{0000--0000/00\$00.00~\copyright~2021 IEEE}
% Remember, if you use this you must call \IEEEpubidadjcol in the second
% column for its text to clear the IEEEpubid mark.

\maketitle

\begin{abstract}
In this article for the first time we applied the eclipse Z-scan technique which is by one order of magnitude more sensitive than the conventional Z-scan technique to measure LiNbO$_3$ crystal's nonlinear refractive index in the THz range. The obtained value of LiNbO$_3$ nonlinear refractive index is estimated to be 5$\pm$2$\times$10$^{-11}$cm$^2$/W which is commensurate with known results. This value correlates with the theoretically calculated nonlinear refractive index coefficient of vibrational nature. The influence of thermal nonlinearity on the experimental results can be neglected, since the estimated temperature induced refractive index change $\Delta$n$_{T}$ equals to 2.6$\times$10$^{-5}$, while addition $\Delta$n$_{nl}$ from optical nonlinearity is 2.9$\times$10$^{-3}$. The demonstrated heightened sensitivity of eclipse Z-scan allows to hold promise for the properties evaluation of materials exhibiting lower nonlinear refractive indices, thus expanding its applicability in characterizing diverse nonlinear optical materials.
\end{abstract}


\section{\label{sec:intro}Introduction}
\IEEEPARstart{N}{onlinear} characteristics of materials such as nonlinear refractive index ($n_2$) and absorption coefficients in the terahertz (THz) frequency range have already been experimentally and theoretically investigated for a number of liquids \cite{tcypkin2021giant,tcypkin2019high, novelli2020nonlinear, francis2020terahertz, guo2020research}, vapors \cite{rasekh2021terahertz, francis2020terahertz, johnson2008water} and crystals \cite{dolgaleva2015prediction, sowade2010nonlinear, zibod2023strong, zhukova2020estimations, jazbinsek2019organic, woldegeorgis2018thz, guselnikov2023, LNB2023}. Within the existing amount of work, the experimental method for determining the value of $n_2$ is used along with theoretical approach \cite{tcypkin2019high,tcypkin2021giant,amini2021regenerative} to prove the validity of the theory \cite{nabilkova2023controlling} and stimulate the development of fundamental research for the evaluation of the coefficient $n_2$ \cite{hemmatian2023simplified, dolgaleva2015prediction}.

Besides, various experimental techniques are used to investigate nonlinear properties of materials in the THz range, each with its unique setup and specific objectives. Examples of such techniques include: Z-scan \cite{sheik1990sensitive,van2018z,melnik2019methodical,he2006direct,10308712}, widely employed for nonlinearities extraction across all spectral ranges, 4f-system Z-scan \cite{Boudebs13,li2024characterizing,wang2018measurement}, pump-probe with time delay \cite{zibod2023strong, nabilkova2022influence, zhao2020ultrafast, hoffmann2009terahertz, sarbak2017direct}, four wave mixing \cite{hosseini2012analysis}, nonlinear ellipse rotation \cite{Miguez14,miguez2017measurement}, and full-phase analysis \cite{francis2020terahertz}. These techniques primarily leverage self-induced lensing phenomena to explore and quantify the nonlinear properties of materials.

It is important to note the orders of magnitude of previously measured and estimated nonlinear refractive indices. For liquids, these values are of the order of 10$^{-10}-$10$^{-9} $cm$^2$/W \cite{tcypkin2021giant, Novelli:24, 10308712}, for solids these values vary from 10$^{-13} $cm$^2$/W to 10$^{-11} $cm$^2$/W \cite{Cornet:14, photonics7040098, zibod2022extremely}. In particular, for lithium niobate (LiNbO$_3$) crystal the value of $n_2$ in the terahertz region is deduced to be 10$^{-11} $cm$^2$/W \cite{Korpa_2016,artser2022radiation}.

The described nonlinearities for solids are couple orders of magnitude lower than for liquids. The sensitivity of the conventional laboratory tabletop Z-scan setup \cite{tcypkin2019high} becomes insufficient for the comprehensive investigation of the nonlinearities that possessed by solids. In particular, the robust interferometric sensitivity characteristic of the Z-scan method, which facilitates the determination of alterations in transmittance due to phase change, emanates from the interference (diffraction) effects engendered by distinct segments of the spatial profile within the far field \cite{sheik1990sensitive}.
    
Only a few unique techniques, such as one consisting of single-mode Fabry-Pérot microcavity with detection by phase contrast imaging technique using a 4f system \cite{LNB2023} or high-powered sources like FEL \cite{kramer2020enabling}, are suitable for evaluating nonlinear refractive index $n_2$ of LiNbO$_3$ in the THz range. Another technique is based on the phenomenon of third harmonic generation disappearing during single-cycle THz pulse propagation in cubic nonlinear medium. It is replaced by the generation of radiation at quadruple frequencies and by the position of the dip in the region of triple frequencies and the peak of quadruple frequency it is possible to evaluate the nonlinear refractive index of LiNbO$_3$ \cite{artser2022radiation}.

One of the alternative method to increase sensitivity of the conventional Z-scan measurements is the eclipse Z-scan \cite{xia1994eclipsing, gomes2007thermally}. The system's sensitivity is enhanced by selectively registering signal counts within the beam peripheries. This is achieved through the obstruction of the central region due to implying disk aperture that mask the central segment of the laser beam. The higher sensitivity to transmission variations derives from the manner light bypasses the obstructing disk, thereby enabling the detection of an "eclipse" and facilitating assessments of the third-order optical nonlinearity inherent in materials. Notably, the application of the eclipse Z-scan method in the context of the THz frequencies represents a novel endeavor, necessitating adaptations to accommodate this spectral domain.

In this work we introduce an advancement in measuring the third-order nonlinearity in the THz spectral range with eclipse Z-scan method. With a one-order sensitivity enhancement over conventional Z-scan techniques, this approach has yielded a remarkable LiNbO$_3$ nonlinear refractive index value of 5$\pm$2$\times$10$^{-11} $cm$^2$/W. The observed value of the refractive index change induced by temperature variations ($\Delta$n$_{T}$), estimated as 2.6$\times$10$^{-5}$, and the dominant refractive index change ($\Delta$n$_{nl}$) attributed to optical nonlinearity, measured as 2.9$\times$10$^{-3}$, assume that the observed nonlinear behavior cannot arise from thermal effects. It could be stated that the nonlinearity obtained has a vibrational nature, since the value obtained coincides with the value theoretically calculated earlier \cite{tcypkin2021giant,photonics7040098} according to the analytical model \cite{dolgaleva2015prediction}. In addition, a comparative analysis of the sensitivities of the conventional and eclipse Z-scan techniques is presented, offering insights into their capabilities.

\section{\label{sec:exp}Experiment}
With a view to reach an order-of-magnitude enhancement of the signal-to-noise ratio of the conventional Z-scan technique, the disk \cite{ferreira2021characterization, gomes2007thermally} instead of the aperture is implemented into setup according to pioneering works in mentioned direction \cite{xia1994eclipsing, sheik1990sensitive}. The application of the disk allows using both open and closed aperture geometry in the same way as aperture does. Here, the nonlinear refractive index $n_2$ of lithium niobate is experimentally determined utilizing the eclipse Z-scan technique in conjunction with a pulsed THz radiation source. 

\begin{figure}[h!]
\centering
\includegraphics[width=1\columnwidth]{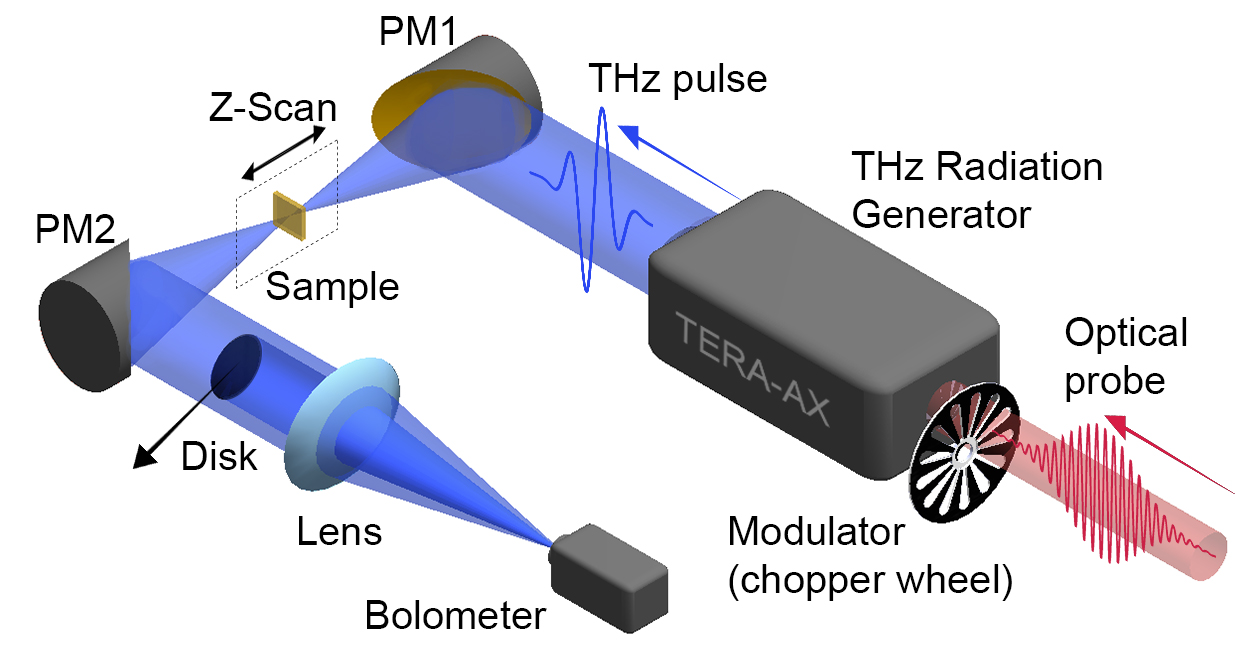}
\caption{
\label{FIG1}
Experimental setup for the eclipse Z-scan measurements. The THz pulse is produced through optical rectification within a MgO:LiNbO$_3$ crystal (Tera-AX Avesta project \cite{F_l_p_2012}), sourced from a 35 femtosecond duration pulse with a central wavelength of 790 nm and a pulse energy of 1 mJ originating from a Ti:Sapphire femtosecond laser with a regenerative amplifier (Regulus Avesta project). Parabolic mirrors (PM1, PM2) are used to focus and collimate the THz pulse. The modulator is used to synchronize the bolometer and the THz radiation source. The disk can be moved from the beam to the open position. A bolometer (Gentec-EO THZ5B-BL-DZ-D0) is used to measure the power of THz radiation.
}
\end{figure}

The experimental setup is illustrated in Fig. \ref{FIG1}. The THz source operates at a repetition rate of 1 kHz. The vertically polarized THz pulse displays a duration of 1 ps, a pulse energy of 400 nJ (with a peak electric field strength of 0.27 MV/cm), and a spectral range extending from 0.1 to 2.5 THz \cite{tcypkin2021giant}. The full width at half maximum (FWHM) of the collimated THz beam is determined using the knife-edge method, yielding values of 8.5 mm for the x-axis and 5.6 mm for the y-axis, as reported in \cite{tcypkin2021giant}. The beam is tightly focused onto the sample via the PM1 mirror (the focal length of the parabolic PM1 mirror is 12.7 mm), resulting in a beam diameter of approximately 1.5 mm at the focal point. The beam diameter values are verified by the knife-edge method with a precision of 100 $\mu$m. The peak intensity at focus reaches 0.5$\times10^8$ W/cm$^2$.  Then the THz radiation is collimated by the parabolic mirror PM2 and focused by the lens with focal length of 50 mm. The slight ellipticity of the beam observed in the experiment should not impact the closed aperture Z-scan traces \cite{mian1996effects}. The metal disk is installed in front of the lens. The diameter of disk is selected relative to the required sensitivity of the eclipse Z-scan technique. Detection is performed using a bolometer (Gentec-EO THZ5B-BL-DZ-D0).

\section{Results}
\subsection{Experimental data of eclipse Z-scan}
\begin{figure}
\centering\includegraphics[width=1\columnwidth]{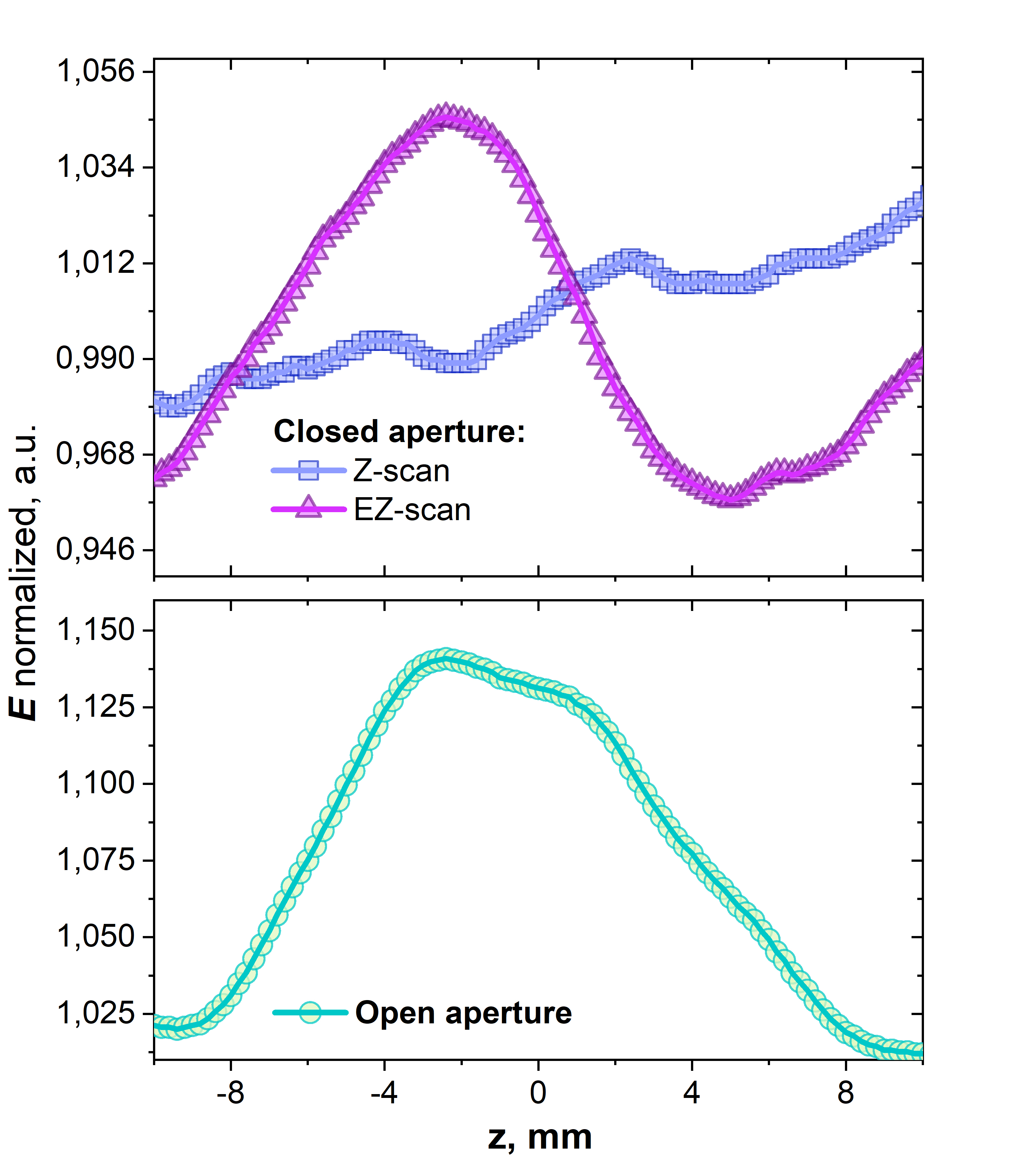}
\caption{
\label{FIG2}
(a) Comparison of the experimental results from closed aperture measurement of Z-scan and eclipse Z-scan (EZ-scan). All obtained curves are normalized to the corresponding mean value to show the difference between $\Delta T$ and order of sensitivity boosted.(b) Experimental results for the case of open aperture measurements without disk.}
\end{figure}

In order to conduct a comparative analysis of the experimental curves obtained with the eclipse Z-scan and conventional Z-scan methods, the 10\% transmittance disk was replaced with a 2\% transmittance aperture. Fig. \ref{FIG2} (a) shows unnormalized closed aperture Z-scan and eclipse Z-scan (EZ-scan) curves obtained experimentally. Fig \ref{FIG2} (b) presents results of the open aperture measurements, which were carried out without disk to exclude the influence of the material's absorption properties. It can be seen that the amplitude of the Z-scan curve fluctuations, where the peak and valley should be, is comparable to the noise of the curves for the EZ-scan. Notably, the difference between the peak to valley in the field \textbf{\textit{E}} registered here for the EZ-scan curves is more than 10 times greater than for the conventional Z-scan curves. 

Fig \ref{FIG3} shows the experimental results of the closed aperture eclipse Z-scan (EZ-scan) measurements normalized to open aperture measurements, compared to the analytical modeling of the EZ-scan. As can be seen in Fig. \ref{FIG2} (a) and \ref{FIG3}, the ends of the curve point upward may be an analytical artifact arising from dividing the closed aperture curve by the open aperture curve due to the presence of noise in both curves.

These experimental results demonstrate that the difference between the normalized peak and valley transmittance $\Delta T$ is found to be 0.021. Thus, using the empirical linear relationship between $\Delta T$ and nonlinear phase shift $\Delta \Phi_0$ \cite{xia1994eclipsing,gomes2007thermally} we can extract the nonlinear refractive index $n_2$ of LiNbO$_3$ crystal by following equation:
\begin{equation}
n_2 = {\frac{\Delta T}{0.68I_{in}}\times \frac{(1-S)^{0.44}}{k L_\alpha }} 
\label{eq1}
\end{equation}

Substituting the experimentally obtained value of $\Delta T$ we found $n_2$ =  5$\pm$2$\times$10$^{-11}$ cm$^2$/W.

\subsection{\label{sec:analytics}Theoretical model of eclipse Z-scan}

The convenience and precision of applying mathematical expressions \cite{sheik1990sensitive} for the THz range and pulsed few-cycle radiation have already been shown \cite{tcypkin2019high, tcypkin2021giant, melnik2019methodical}. In our simulations, the field distribution $E$ assuming a TEM$_{00}$ Gaussian beam of waist radius $w_o$ traveling in the $+z$ direction is given by:
\begin{equation}
E(r,t,z)=E_0(t) \frac{w_0}{w(z)} \exp\left({\frac{-r^2}{w(z)^2}}-{\frac{ikr^2}{2R(z)}}\right) e^{-i\phi(r,t,z)}
\label{eq2}
\end{equation}
where $E_0(t)$ is the initial temporal envelope of the input field $E_0 \sin(2\pi \nu_0 t)$, $w(z)^2 = w^2_0(1 + z^2/z^2_0)$ is the beam radius, $R(z) = z(1 + z^2/z^2_0)$ is the radius of curvature of the wave front at $Z$, $z_0 = kw^2_0/2$ is the diffraction length of the beam, and $k = 2\pi/\lambda$ is the wave vector. The $e^{-i\phi(z,t)}$ term contains all the radically uniform phase variations. In the thin sample slowly varying envelope approximation, the electric field amplitude only changes due to the absorption, while the phase of the electric field due to the nonlinear refraction resulting in nonlinear phase term $e^{-i\Delta\phi(r,t,z)}$. Gaussian decomposition (GD) method is implemented according to \cite{sheik1990sensitive} to calculate the complex field distribution $E_a$ in the aperture plane. So $E_a$ is derived as:
\begin{equation}
\begin{gathered}
E_a(r,t,z)=E(z,t)\exp(-\alpha L/2)
\times \\
\times \sum_{m=0}^{+\infty} {\frac{[-i \Delta \phi_0 (z,t)]^m}{m!}}{\frac{w_{m0}}{w_m}}\exp({\frac{-r^2}{w_m^2}}-{\frac{ikr^2}{2R_m}}+iQ_m)
\label{eq3}
\end{gathered}
\end{equation}
where $\Delta \phi_0(z,t)=\Delta \Phi_{0}(t)/(1+z^{2}/z_{0}^{2}), \Delta \Phi_0(t)=k \Delta n_0(t)L$ is the nonlinear phase shift. The GD method involves decomposing of the electric field at the sample exit plane into Gaussian beams using a Taylor series expansion of the nonlinear phase term. This approach is based on the assumption of small phase changes, allowing to take into consideration only the initial terms of the expansion. Therefore, defining $d$ as the propagation distance in free space from the sample to the disk, and $g=1+d / R(z)$, other parameters in Eq. \ref{eq3} according to \cite{sheik1990sensitive} are expressed as:

\begin{figure}
\centering\includegraphics[width=1\columnwidth]{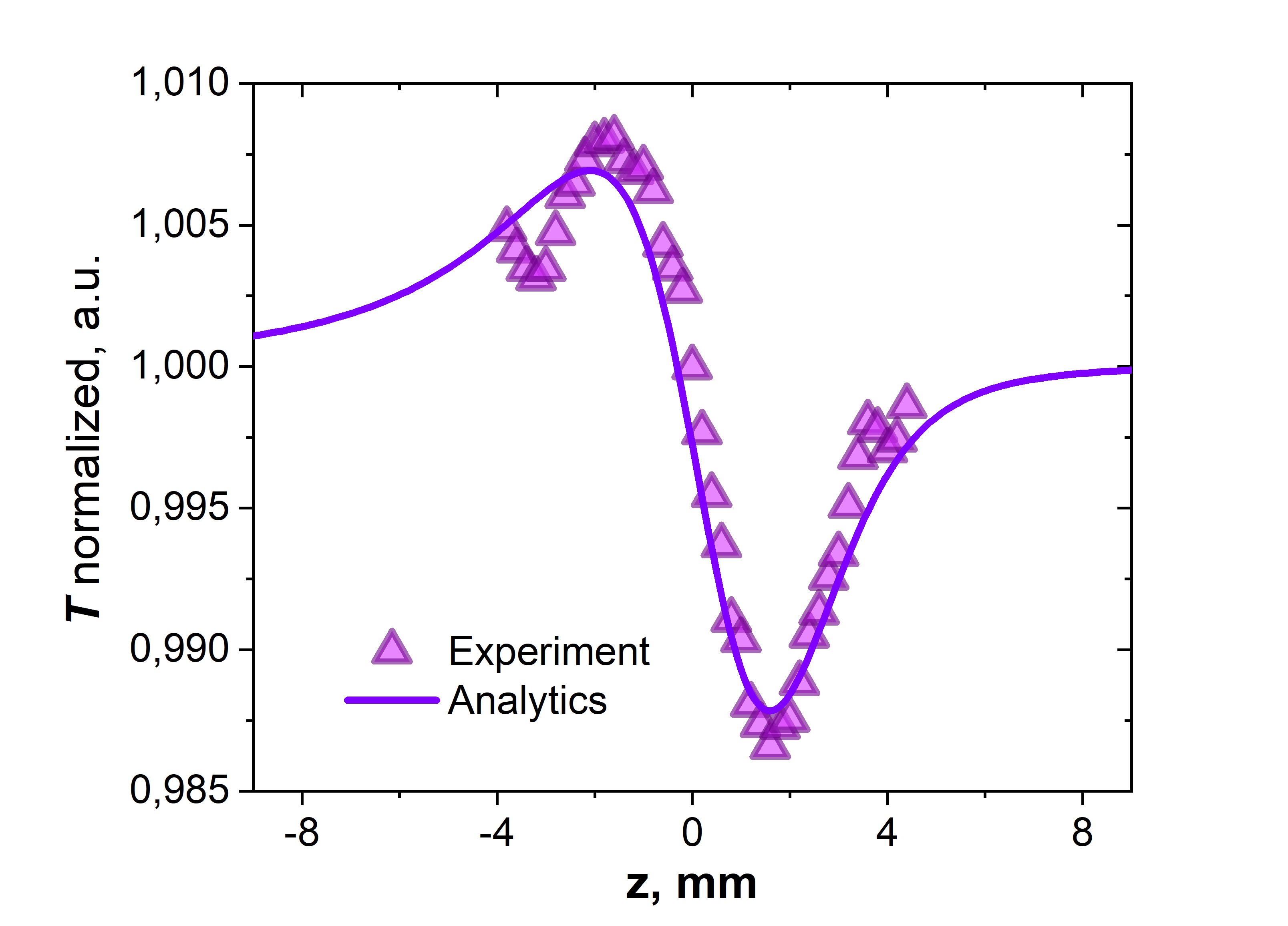}
\caption{
\label{FIG3}
Comparison of the experimental results of the Z-scan method for the pulsed broadband THz radiation for the LiNbO$_3$ crystal with an analytical Z-scan curve. The analytical curve is calculated using Eq. \ref{eq10}.}
\end{figure}

\begin{gather}
w_{m 0}^{2}=w_{z}^{2} /(2 m+1)  \\
d_{m}=k w_{m 0}^{2} / 2  \\
w_{m}^{2}=w_{m 0}^{2}\left(g^{2}+d^{2} / d_{m}^{2}\right) \\
R_{m}=d\left(1-\frac{g}{g^{2}+d^{2} / d_{m}^{2}}\right)^{-1}  \\
\theta_{m}=\tan ^{-1}\left(\frac{d / d_{m}}{g}\right) 
\end{gather}
The transmitted through the disk power is obtained 
by spatially integrating $E_a(r,t,z)$ from the disk radius 
$r_d$ to the beam radius on the aperture $w_a$ giving:
\begin{equation}
P_{T}(\Delta \Phi_{0}(t))=c\epsilon_0 N_0 \pi \int_{r_d}^{w_a}|E_a(r,t)|^2rdr
\label{eq9}
\end{equation}
Integrating up to beam radius on the aperture $w_a$ allows neglecting the impact of "noise", where the beam profile fluctuates around zero. Thus, the normalized transmittance can be calculated as
\begin{equation}
T(z) = {\frac{\int_{-\infty}^{+\infty}P_{T}(\Delta \Phi_{0}(t))dt}{S\int_{-\infty}^{+\infty}P_{i}(t)dt}}
\label{eq10}
\end{equation}
where $P_i$(t)=${\pi w_{0}^2 I_0(t)/2}$ is the instantaneous input power (within the sample), $S$=1-$\exp(-2{r_{d}^2/w_{a}^2})$ is the disk linear transmittance.

\begin{figure}[h!]
\centering\includegraphics[width=1\columnwidth]{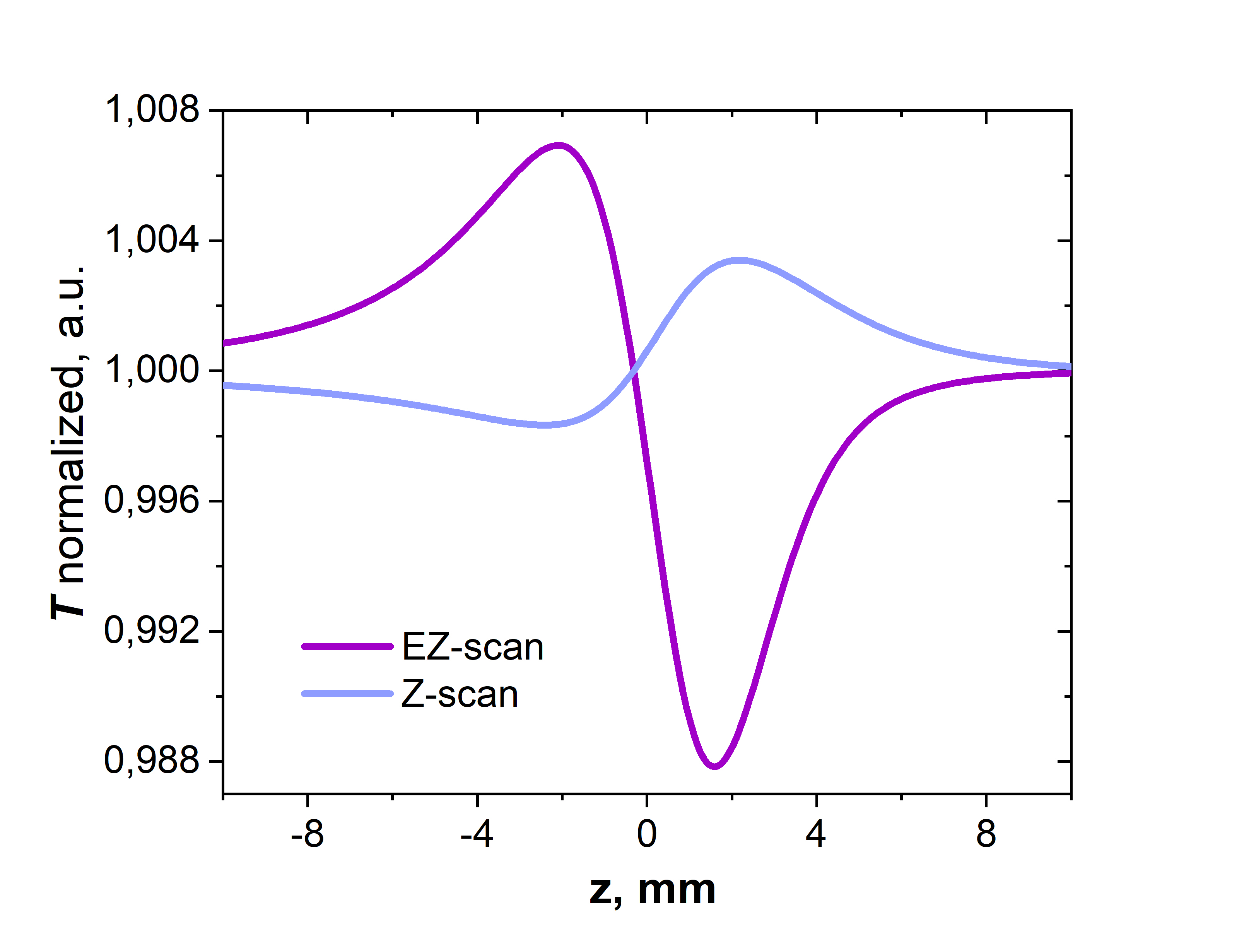}
\caption{
\label{FIG4}
Comparison of the analytical results of the Z-scan and eclipse Z-scan (EZ-scan) modeling.}
\end{figure}
By gathering the setup parameters and the beam characteristics from the experiment, the nonlinear refractive index $n_2$ value calculated from \ref{eq1}, and substituting them into equation \ref{eq10} (and nested equations), we can obtain the normalized transmittance \textit{T} from the coordinates of crystal displacement \textit{z}. Fig. \ref{FIG4} shows a comparison of analytical eclipse and conventional Z-scan curves, demonstrating the order of sensitivity enhancement for the former method. The Z-scan curves are obtained with the same theoretical model, however Eq. \ref{eq9} is adjusted to perform integration over the aperture: from 0 to $r_a$.

\section{Discussion}
The experimental as well as the analytical findings underscore the efficiency of employing eclipse Z-scan instead of conventional Z-scan method, culminating in an enhancement in the sensitivity of THz systems by an order of magnitude. The achieved order of nonlinearity, approximately 10$^{-11}$ cm$^2$/W, stands as a noteworthy accomplishment. As was mentioned, the diameter of the metal disk, which determines the parameter \textit{S}, was chosen on the basis of two considerations: the sensitivity of the eclipse Z-scan method and the sensitivity of the detector. As a result, a disk with a diameter that covers 90\% of the incident radiation was chosen, which allowed us to perform measurements with a signal-to-noise ratio of 10 and to obtain a significant drop in the Z-scan curve.

Certain types of light sources can induce cumulative interactions, such as pulsed THz systems with large pulse durations or high repetition rates. In such systems, thermal effects may occur. In the context of determining the nonlinear refractive index of liquids \cite{tcypkin2019high, tcypkin2021giant, novelli2020nonlinear}, temperature-induced nonlinear effects were deemed negligible, as each pulse interact with a refreshed surface of liquid jet. However, for crystals, consideration of temperature fluctuations during the experiment becomes mandatory for a comprehensive evaluation of their influence on the refractive index.
Strong thermal effects can contribute to the refractive index by inducing temperature fluctuations through non-radiative interactions within the system. The upper bound evaluation was made for temperature nonlinear refractive index contribution $n_{2_T}$ in comparison with the contribution of vibrational nonlinear refractive index $n_{2_{nl}}$ to the change in the refractive index $\Delta n$.

To estimate the thermal effects per laser repetition rate, we use the following expression $n_{2_{th}} = \frac{dn}{dT}\frac{\alpha R^2}{\kappa}$ \cite{boyd2008ultrafast}. Here $\frac{dn}{dT}$ displays the speed of refractive index change with temperature variation, it is found to be 7.03$\times$10$^{-4}$ 1/K \cite{sowade2010nonlinear}. $\alpha$ = 9 cm$^{-1}$ denotes the linear absorption coefficient of the material, $\kappa$ = 0.046 W/(cm$\times$K) stands for the thermal conductivity, and $R$ = 0.7 mm is a radius of a circular laser beam in focus.

The response time $\tau_r$ for the thermal nonlinearity is given by $\tau_r \approx \frac{\rho C R^2}{\kappa}$ \cite{Ikeda:08}, so thermally induced $\Delta n$ per one repetition of pulses would be $\frac{n_{2~th} I \tau_p }{\tau_r}$, because interaction of the crystal with the beam occurs for about pulse duration $\tau_p$, which is equal to 1 ps. Here, material's density $\rho$ is 4700 kg/m$^3$, heat capacity $C$ is 628 J/(kg K).

The energy per one repetition can be estimated as $Q = \epsilon P m$, where pulse energy $P$ is equal to 400 nJ, emissivity of the crystal $\epsilon$ is 0.6, and number of pulses per one repetition $m$, controlled by chopper's speed, is 40. If we assume that all obtained energy is used for heating, then the crystal heats up according to the equation $\Delta T_h= \frac{Q}{C m}$, where $m =\rho \pi R^2 L $ is mass of illuminated area of the crystal. Then, during the time when the chopper closes the laser, the crystal cools by an amount $\Delta T _c = \frac{\tau h S \Delta T_h}{m C}$, where $S$ is an illuminated area and $h$ is a heat transfer coefficient equal to 10 W/(m$^2$ K). Accordingly, the temperature $\Delta T$ to which the crystal is heated during one passage will be equal to $\Delta T_h$ - $\Delta T _c$. The overall heating temperature of the sample for the entire duration of the experiment is 0.1$\times$ 10$^{-3}$ K. 

As an upper estimate, we assume that the temperature change $\Delta T$ is constant while the spot size of the laser beam on the crystal should change as the sample is displaced through the beam focus in the $z$ direction. The sample passes this path in 10 seconds, so the crystal has time to interact with radiation and heat up by $\Delta T$ in 250 repetitions, according to repetition rate of laser and frequency of chopper. Therefore, accumulative thermal contribution to refractive index change  $\Delta n_{th}$ is 2.6$\times$ 10$^{-5}$, while contribution of vibrational nonlinearity $\Delta n_{nl}$ is 2.9$\times$ 10$^{-3}$. 

\section{Conclusion}
This study shed a light on an adjustment to tried and tested in THz range Z-scan technique, that offer to gain sensitivity of experiment. Eclipse Z-scan methodology was used for assessing third-order nonlinearity within the THz spectral domain. Demonstrating a sensitivity enhancement of one order compared to conventional Z-scan methodologies \cite{tcypkin2019high, tcypkin2021giant}, our refinement has unveiled a noteworthy nonlinear refractive index value $n_2$ for LiNbO$_3$ crystal of 5$\pm$2$\times$10$^{-11} $cm$^2$/W. Additionally, the theoretical approach, that describes the behavior of light in a crystal when conducting eclipse Z-scan measurement, was modified to justify the obtained value of $n_2$. Our results show that experimentally obtained Z-scan measurement curves and simulations of self-focusing processes in LiNbO$_3$ crystal are in good agreement. Notably, our investigation into temperature-induced variations in refractive index ($\Delta$n$_{T}$), estimated as 2.6$\times$10$^{-5}$, alongside the predominant refractive index change ($\Delta$n$_{nl}$) attributed to optical nonlinearity, measured at 2.9$\times$10$^{-3}$, suggests that the observed nonlinear phenomena mostly ascribed to optical nature. This assertion is supported by the congruence between our measured nonlinearity and previously theorized values \cite{tcypkin2021giant,photonics7040098}, as predicted by the analytical framework proposed by \cite{dolgaleva2015prediction}, indicating a vibrational underpinning to the observed nonlinear behavior. Furthermore, a comparative evaluation of the sensitivities inherent to conventional and eclipse Z-scan methods is provided experimentally and analytically, affording valuable insights into their respective efficiency and potential applications. 

The results presented emphasize the importance of exploring the nonlinear refractive index of diverse media to identify materials exhibiting high nonlinearity, thereby facilitating the advancement of nonlinear terahertz optics. The nonlinear refractive index plays a pivotal role in determining the optical response of materials, enabling the manipulation of light for various applications in photonics and optoelectronics. Devices such as semiconductor optical amplifiers, flip-flops, switchers, optical transistors, and modulators have been identified as examples of systems with high nonlinear refractive index, showcasing the practical relevance of investigating this property. By exploring and harnessing the nonlinear refractive index of materials, researchers can unlock new avenues for enhancing the performance and functionality of optical devices. This research not only contributes to expanding our fundamental understanding of light-matter interactions but also holds immense promise for driving innovation in the field of nonlinear optics, for example, developing devices for creating all-optical computer.

\section*{Acknowledgments}
All authors acknowledge support from the Russian Science Foundation (Grant No. 24-22-00084).

\bibliographystyle{IEEEtran}
\bibliography{IEEE_ECLIPSE_BIBL}

%\begin{thebibliography}{1}
%\bibliographystyle{IEEEtran}

%\bibitem{ref1}
%{\it{Mathematics Into Type}}. American Mathematical Society. [Online]. Available: https://www.ams.org/arc/styleguide/mit-2.pdf
%\end{thebibliography}

\end{document}